\documentclass[aps,twocolumn,nofootinbib]{revtex4}
\usepackage{amsfonts}
\usepackage{amssymb}
\usepackage[dvips]{graphics}
\usepackage[brazil]{babel}
\usepackage[latin1]{inputenc}

\newcommand{\be}{\begin{equation}}
\newcommand{\ee}{\end{equation}}
\newcommand{\bea}{\begin{eqnarray}}
\newcommand{\eea}{\end{eqnarray}}
\newcommand{\beann}{\begin{eqnarray*}}
\newcommand{\eeann}{\end{eqnarray*}}

\begin{document}

\author{ E. Abdalla and M. S. Hussein}
\email{eabdalla@if.usp.br, hussein@if.usp.br}
\affiliation{Instituto de Física, Universidade de São Paulo\\
C. P. 66318,
05315-970 S\~ao Paulo, Brasil }
\title{
{\bf Ultra cold neutrons: determination of the electric dipole moment
and gravitational corrections via matter wave interferometry}}
\date{\today}

\begin{abstract}
We propose experiments using ultra cold neutrons which can be used to
determine the electric dipole moment of the neutron itself, a well as to test
corrections to gravity as they are foreseen by string theories and
Kaluza-Klein mechanisms. 
\end{abstract}
\maketitle

Matter Wave Interferometry Devices (MWID) are aimed at testing one of
the important pilars of quantum mechanics, namely wave-like interference 
of matter. These devices use the idea that quantum mechanical 
interference results in diffraction patterns of the intensity, which 
critically depend on the relative phase of the interfering waves. This 
allows the extraction of useful information about the phase generating, 
refractive-diffractive medium through which the wave propagates.

There are two main questions we are concerned with here, which appear in
the context of grand unification and superstring theories. The first
is that within grand unified theories with supersymmetry
there are several plausible contributions to the electric dipole
moments of eletrons and neutrons \cite{Keumkong,nir}, a problem
known as the SUSY CP problem. CP violation is very little known, and
the violating parameters, especially the neutron and the eletron
electric dipole moments should be tested.

In fact, the Standard Model (SM) is an extremely efficient theory in what
concerns experience today, but it is widely believed that there exists new
physics beyond it and any extension of the SM, especially in the case it
contains supersymmetry (SSM), which is also widely believed to be certain,
foresees CP violation, 
since it is necessary to explain, from the observational point of view, the
baryon asymmetry of the Universe \cite{baryonasymmetry}.

In the case of the neutron electric dipole moment the relevant
CP-violating terms are flavour blind. The contribution from the
Kobayashi-Maskawa matrix arises at three loop level, and the experimental
bound is \cite{harris}
\be
d_n \le 6.3\times 10^{-26}e\times cm\quad .
\ee

The second question we deal with here concerns a specific point of string
theories, which is the fact that they require the existence of further
dimensions \cite{gsw,polchinski,scherk} which can be both of infinite
\cite{horavawitten,randallsundrum,abdetall} or finite
\cite{arkanihamed,randallsundrum1} size. This  leads to corrections of
Newton's gravity law at very small distances. 

An experimental confirmation (or denial) of the above facts lies in
the heart of the question whether grand unified theories with
supersymmetry and string theories are part of physics or are just
mathematically well defined objects beyond physical reality.

In a recent publication \cite{1}, an electric field-induced phase was
measured with great precision for atomic waves. We propose that the same
idea may be used to extract the dipole moment of the neutron, and test the
behaviour of gravity for very short distances.

The first idea is to use Ultra-Cold Neutrons (UCN) of a velocity of 
$v_n \simeq 5m/s$, which 
pass through a splitter (crystal) before they actually interact with
the electric field of an appropriate capacitor. The idea of using
ultracold neutrons in order to probe properties of elementary particles
has already been advocated \cite{orfeu}. When a neutron moves
in a potential, $U(x)$, the phase of its wave function can be written
as (considering this potential much smaller than its kinetic energy)
\be\label{eq1}
\Delta \phi \approx - \frac{1}{\hbar v_n} \int U(x)dx\quad ,
\ee
where the limits of the integral are determined by the length of the
phase generating apparatus. If $U(x)$ is a constant, then 
\be\label{eq2}
\Delta \phi \approx - \frac{1}{\hslash v_n} U L\quad . 
\ee

Therefore, the interference of the two split neutron waves, one of
which affected by (\ref{eq2}), would behave as
\be\label{eq3}
I_n  \approx  A + B cos(\phi_0 + \Delta\phi cos\theta)\quad ,
\ee
where $\theta$ is the angle between the parallel to the
propagation-in-medium wave, and that of the ¨screen¨.

As the neutron wave traverses the region of strong electric field it
feels an interaction which has the form
\be\label{eq4}
U = \vec{d_n} \cdot \vec{\varepsilon} - \mu_n  \left(\frac{v_n}{c}
\varepsilon\right) - \frac{\alpha_n \varepsilon^2}{2}\quad ,
\ee
where the term $\left(-\frac{v_n}{c}\varepsilon\right)$ is the induced
magnetic field owing to the motion of the particle through the electric
field $\varepsilon$. The third term corresponds to the neutron stark
effect interaction where the electric polarizability $\alpha_n$ enters
into the picture. Of course, at the very slow velocities we are
considering, gravity effects are relevant. These, however can be
eliminated by a proper calibration of the experiment through a
measurement the matter interference when the electric field is switched off.

\begin{table}[t!]
\begin{center}
\begin{tabular}{|c|c|}\hline
\multicolumn{2}{|c|}{{\bf Table: fundamental constants }}\\ \hline
Neutron mass & $m_n= 1.67 \times 10^{-27}$Kg\\ \hline 
Magnetic dipole moment& $-1.91 \times \frac{e\hbar}{2m_pc^2}$ \\ \hline 
Neutron Lifetime&      $\tau = (889.1\pm2.1)s$ \\ \hline 
Electric polarizability&    $\alpha = (0.98\pm 0.20)\times 10^{-42}$cm$^3$
\\ \hline  
Charge&    $ (-0.4\pm 1.1)\times 10^{-21}e $      \\ \hline 
Electric Dipole moment&    $d_n \le 6.3\times 10^{-26}e\times cm$    \\ \hline 
Newton constant& $G= 6.67 \times 10^{-11} $m$^3$/Kg s$^2$   \\ \hline 
\end{tabular} 
\end{center}
\end{table}

From the numbers shown in the table we anticipate that the third term in
Eq. (\ref{eq4}) is much smaller than the first two, even for the reasonably
strong field considered, namely $e \varepsilon \sim \frac{100KeV}
{cm}$. In so far as the competition between the electric and 
magnetic interaction of the neutron is concerned, we have evaluated both 
terms using the values of $\mu_n$ and $d_n$ given in the table, a neutron 
velocity of $5m/s$ and the above cited value of $\varepsilon$. We find
\be\label{eq5}
\frac{d_n}{\mu_n \frac{v_n}{c}} \approx 10^{-4} \quad .
\ee

It is clear that the measurement of the phase shift of the two beams
of neutrons will simply be sensitive to $\mu_n$.

In order to remove the effect of the magnetic interaction to first
order we suggest using an external magnetic field that is opposite to
the induced one, 
\be\label{eq6}
\vec{B}_{ext} = \frac{\vec{v}_n}{c} \times  \vec{\varepsilon}\quad . 
\ee

This field is quite weak and can be easily attained. In fact, due to
the smallness of (\ref{eq5}), there is no practical way of cancelling such a
term with the corresponding one in (\ref{eq4}) to the desired accuracy, since
the velocity $v_n$ itself is defined to an accuracy of a few percent. 
However, if we allow for several total reflections, $N$, of the neutrons 
from mirrors placed at the ends of the capacitor this would on the one 
hand enhance the electric interaction as the phase becomes 
\be\label{eq7}
\Delta\phi_e =  - \frac{NUL}{\hbar v_n} = \frac{N}{\hbar v_n}  d_n (eV)
\frac{L}{a}\quad ,
\ee
with a still very small (almost cancelled) magnetic interaction which
reverses sign for a neutron traversing the cylinder back and forth.
In eq. (\ref{eq7}), $eV$ is the voltage across the capacitor and $a$ is
the distance between the plattes. Taking $\frac{L}{\alpha} = 10^4$ ($L =
1$m, $a = 1$mm) and using for $N \simeq 1001$, (odd number of collisions
guarantees that the neutrons come out at the other end of the capacitor),
we find (for $d_n \sim 0.6\times 10^{-25}$e$\,$cm)
\be\label{eq8}
\Delta\phi_e  \simeq 2\times 10^{-3} rad \simeq 0.1^o  \quad ,
\ee
a measurable shift in the diffraction maxima seen on the screen. For
$N\sim 1000$ and an external $B$ field given with an accuracy of 1\% the
difficulty related to the smallness of the electric dipole moment,
(\ref{eq5}) is overcome (with the cancellation of the velocity dependent
term for back and forth travels) and we have a result with a 10\%
confidence. In fact
angular shifts of the order of $10^{-3}$ have already been measured in
another quantal wave mechanical problem involving Mott-type oscillation in
the elastic scattering of two slowleads ions \cite{1}. The
purpose of these measurements was the verification of the possible
existence of long-range multigluon exchange interaction (colour van der
Waals force) which were predicted to exist in several theories
\cite{husseinetal0}. Therefore an enhancement of one order of magnitude
in the neutron electric dipole moment limit can be obtained.

We should emphasize that, in $\Delta\phi_e$, we have na very small
number, $d_n (\simeq 10^{-25} e cm)$ multiplying a very large number, $N
\left(\frac{L}{a}\right) \frac{1}{\hbar v_n}$. This latter can be
made larger with a suitable changes in the macroscopic variables
that define it. For a time of flight approximately equal to the neutron
lifetime, $N$ can be as large as 5000.

We now pass to the second proposal. As mentioned, according to string
theory, world is multidimensional. The question of how do the extra
dimensions behave is however not fixed, at least from the theoretical
side, since there are several possibilities, namely the extra dimensions
might be of the size of the Planck lenght \cite{gsw}, they can have a
finite submilimetric size in order to unify the Planck and Standard Model
scales \cite{arkanihamed} or they can be infinite with a nontrivial warp
in the extra dimensions \cite{randallsundrum}. It is an important
experimental problem to prove these extra dimensions by proposing new
situations where they can leave an experimental imprint
\cite{abdetall,husseinetal}. A recent proposal in this direction was given
in \cite{newhussein}.

We thus propose to use the slow neutrons to measure the new
contributions to Newtons potential. We suppose that
since gravitational fields are weak, the metric, either
$3+1$-dimensional, or $4+1$-dimensional is well described in terms of a
potential as $g_{00} = -1 + \frac{2U_g}{c^2}$, with $U_g$ given by the
solution of the Poisson equation.
We propose that the neutrons are colimated into a small hollow
cilinder. In accordance with $3+1$ dimensional Newton law the
potential due to a line of matter with linear density $\lambda$,
located at the $z$-axis from $z = 0$ to $z = L$ is 
\be\label{eq9}
U_g(x) = G \lambda \ln  \frac{(L-z) + \sqrt{r^2 + (L  -
z)^2}}{-z + \sqrt{r^2 + z^2}}  \quad ,
\ee
where $r$ is the distance to the $z$-axis. We shall suppose that
for a hollow cilindes the potential inside the cilinder is given by
the above with $r = r_0$ being the internal radius.

For a 4+1-dimensional world the expression is different, and we
suppose that it is effective only for distances smaller than a typical
size of the extra dimension $\xi \sim 0.1 $mm. Therefore, inside the
hollow cylinder we subtract the 3+1-dimensional contribution of a slice $z
\,\epsilon\, [z -\xi, z + \xi]$ and add a 4+1-dimensional
contribution of a slice $z\, \epsilon \, [z -\xi, z +
\xi]$, obtaining\newpage
\bea
U_g (z)&\simeq& G\lambda\left\{ \ln\frac{(L-z)
+\sqrt{r^2 +(L-z)^2 }} {-z
+\sqrt{r^2 +z^2}} +\frac {2\xi} r  \right.\nonumber\\
&&\qquad\qquad  \left. - \ln\frac{\xi
+\sqrt{r^2 +\xi^2 }} {-\xi
+\sqrt{r^2 +\xi^2}} \right\}\\
&\simeq&  G\lambda \left\{ \ln\frac{4z(L-z)}{r^2} 
+\frac {2\xi} r   -  \ln\frac{\xi +\sqrt{r^2 +\xi^2 }} {-\xi
+\sqrt{r^2 +\xi^2}} \right\}\quad .\nonumber
\eea

The order of magnitude of the effect on a slow neutron is not
negligible, that is
\be\label{eq11}
\Delta \phi_g=N\frac{U_gm_nL}{\hbar v}  \simeq
N\frac{G \lambda m_nL}{ \hbar v} =   6\times 10^{-3} = 0.3^{o}
\ee
for a cylinder of linear density $\lambda \sim 0.03$Kg/m, and $N\sim 1000$,
which is comparable to the value of $\Delta\phi_e$ in eq. (\ref{eq8}) for
the electric dipole measurement. The above numbers are compatible with a
thin cylinder of radius $1$mm. For a cylinder of $0.3$mm, quite
comparable to the presumed size of the extra dimension, we still get a 
measurable result \cite{1}. Notice that  $N\sim 5000$ is still compatible
with the neutron lifetime.

ACKNOWLEDGEMENT: This work was partially supported by FAPESP and CNPQ, Brazil.


\begin{thebibliography}{nn}
\bibitem{Keumkong} Y.Y. Keum and Otto  C.W. Kong, {\it Phys. Rev.}
{\bf  D63} (2001) 113012,  hep-ph/0101113; {\it Phys. Rev. Lett.}
{\bf 86} (2001) 393-396, hep-ph/0004110.
\bibitem{nir} Yosef Nir, 27th SLAC Summer Institute on Particle Physics:
CP Violation in and Beyond the Standard Model (SSI 99), Stanford,
California, 7-16 Jul 1999, *Trieste 1999, Particle physics* 165-243,
hep-ph/9911321.
\bibitem{baryonasymmetry}A. D. Sakharov {\it JETP Lett.} {\bf 5} (1967)
24; G. R. Farrar and M. E. Shaposhnikov {\it Phys. Rev.} {\bf D50} (1994)
774; P. Huet and E. Sather  {\it Phys. Rev.} {\bf D51} (1995) 379.
\bibitem{harris} P.G. Harris et al  {\it Phys. Rev. Lett.} {\bf 82} (1999)
904.
\bibitem{gsw} Michael B. Green, J.H. Schwarz, Edward Witten, 
{\it Superstring Theory}, Cambridge Monographs On
Mathematical Physics, 1987.
\bibitem{polchinski}J. Polchinski {\it String Theory} 
Cambridge Univ. Pr., 1998.
\bibitem{scherk}J. Scherk {\it Rev. Mod. Phys.} {\bf 47} (1975) 123-164.
\bibitem{horavawitten} Petr Horava and Edward Witten,
{\it Nucl. Phys. } {\bf B460} (1996) 506-524,  hep-th/9510209.
\bibitem{randallsundrum} Lisa Randall and Raman Sundrum
{\it Phys. Rev. Lett.} {\bf 83} (1999) 4690-4693.
\bibitem{abdetall}Elcio Abdalla, Adenauer Casali, Bertha Cuadros-Melgar,
  {\it   Nucl. Phys.} {\bf B644} (2002) 201-222,    hep-th/0205203.
\bibitem{arkanihamed} Ignatios Antoniadis, Nima Arkani-Hamed, Savas
Dimopoulos and G.R. Dvali
{\it Phys. Lett.} {\bf B436} (1998) 257-263, hep-ph/9804398.
\bibitem{randallsundrum1} Lisa Randall and Raman Sundrum
{\it Phys. Rev. Lett.} {\bf 83} (1999)3370-3373.
\bibitem{orfeu}O. Bertolami and F. M. Nunes {\it Class. Quantum Grav. }
{\bf 20} (2003) L61-L66.
\bibitem{1}A.C.C. Villari {\it et al} {\it Phys. Rev. Lett.}
{\bf 71} (1993) 2551.
\bibitem{husseinetal0}M. S. Hussein, C. L. Lima, M. P. Pato and
C. A. Bertulani {\it Phys. Rev. Lett.} {\bf 65} (1990) 839.
\bibitem{husseinetal}M. S. Hussein. R. Lichtenth\"aler, M. P. Pato and
C. A. Bertulani {\it Braz. J. Phys.} {\bf 27} (1997).
\bibitem{newhussein}A. Frank, P. Van Isacker and J. Gomez-Camacho
{\it Phys. Lett.} {\bf B582} (2004) 15-20.
\end{thebibliography}
\end{document}